# Intelligent Fault and Lightning Detection Algorithm for VSC-MTDC grids based on ResNet with Hybrid Attention Mechanism

Yunqi Zhang

***Abstract*-- To address the existing challenges in fault detection for voltage source converter-based multi-terminal DC (VSC-MTDC) grids, this paper proposes an intelligent fault and lightning detection algorithm based on S transform and Residual Network with hybrid attention mechanism (RWHAM). The DC line double-ended initial current traveling waves (ICTWs) are first converted into time-frequency matrices by performing S transform and then visualized as a two-dimensional image. The image effectively characterizes the time-frequency features of ICTWs under different fault and lightning conditions, making it easier to extract critical features for the models. Next, the RWHAM model is constructed and the images are fed into the model to detect internal and external faults and lightning interference. The weight of the key information in the image is increased by hybrid attention mechanism, which in turn improves the fault detection ability of the RWHAM model. Extensive simulations involving 21,620 distinct cases on PSCAD/EMTDC validate the algorithm's high accuracy and rapid response across different types of faults and lightning interference. Furthermore, it exhibits excellent generalization ability when the parameters of VSC-MTDC grids change.**

## I. Introduction

THE voltage source converter-based high voltage DC (VSC-HVDC) grid offers flexible power control and low system harmonic content, demonstrating significant advantages in large-scale renewable energy integration, asynchronous grid interconnection, and islanded power supply, leading to increasingly widespread applications [1-3]. For large-scale, long-distance VSC-HVDC transmission systems, the probability of faults and lightning interference occurring on DC lines increases further, making accurate and rapid fault and lightning detection technology crucial for ensuring the safe and stable operation of flexible DC systems.

Currently, the challenges faced by VSC based multi-terminal high voltage DC (VSC-MTDC) systems mainly include the following: On one hand, with the continuous increase in renewable energy capacity, VSC-MTDC grids are expanding in scale, featuring more complex topologies and increasingly prominent power-electronics-dominated characteristics. This leads to greater randomness and nonlinearity in power system faults and lightning interference. Additionally, issues such as measurement errors in electrical signals make fault and lightning interference mechanism analysis based on physical models increasingly difficult, resulting in insufficient accuracy in traditional fault and lightning detection methods. On the other hand, VSC-MTDC grids impose extremely high demands on fault detection speed. When a fault occurs in the VSC-MTDC grid, due to the small damping of the fault loop, the sub-module capacitors of VSC discharge rapidly, and the fault current can reach several tens of times the rated value within a few milliseconds. Given the susceptibility of VSC semiconductor devices to overcurrent damage, DC circuit breakers must operate within milliseconds after fault occurs, leaving very limited transient data available for analysis. Therefore, there is an urgent need to study fault and lightning detection algorithms for VSC-MTDC grids that can simultaneously satisfy the requirements of accuracy and rapidity.

At present, most of the research on HVDC fault detection principles is merely based on the fault mechanism characteristics of the physical model. Additionally, research on lightning interference detection is relatively limited. In reference [4], the fault zone is discriminated based on the amplitude, polarity and symmetry of the double-ended current mutations. However, its reliability and sensitivity are greatly impacted by the fault impedance, and it is only applicable to short lines or long lines equipped with ultra-fast disconnecting breakers. Reference [5] proposes a fault detection scheme based on the second-order difference of backward voltage traveling waves. The required data window is 5ms, which cannot meet the rapidity requirements of the main protection of the VSC-HVDC grid. Reference [6] identifies the faulty line based on the transient high-frequency energy magnitude of the DC line current. This scheme simplifies the DC line into an RL lumped parameter model and the errors caused by wave dispersion and attenuation effects are not taken into account.

This work was supported by the State Grid Corporation of Headquarters Science and Technology Project under Grant 5500-202356406A-2-4-KJ. (Corresponding author: Yunqi Zhang).

Yunqi Zhang, Yue Yu and Guosheng Yang are with State Key Laboratory for Security and Energy Saving, China Electric Power Research Institute, Beijing 100192, China. (email:zhangyunqi@epri.sgcc.com.cn).

The reliability will decrease as the voltage level and length of the transmission line increase. The studies in reference [7] show that the exponents in the theoretical expressions of the voltage traveling waves (VTWs) under internal faults are much larger than external faults, and the internal faults can be determined based on the magnitude of the exponent. Reference [8] indicates that the frequency-domain attenuation rate of initial voltage traveling waves under external faults is higher than that under internal faults, and the internal and external faults are detected based on this rate. Reference [9] calculates the similarity of the time-frequency matrices of the double-ended initial current traveling waves (ICTWs), and identifies the internal and external faults based on the similarity value. Furthermore, reference [10] proposes to identify internal faults and lightning interference based on the ratio of high-frequency and low-frequency energy of ICTW. References [4-10] all conducted analyses on the voltages or currents under internal and external faults for the VSC-HVDC grid. These analyses were based on the equivalent physical model of the VSC-HVDC grid, and the system parameters were simplified to varying degrees, leading to deviations from actual fault transient processes. Furthermore, these methods require setting criterion thresholds manually based on either simulation results or analytical expressions. However, the setting of such thresholds is susceptible to multiple influencing factors, including system impedance parameters, topological structure, fault or lightning types, harmonics and so on. This makes it difficult to define thresholds that are universally applicable to all conditions, ultimately resulting in insufficient accuracy in fault and lightning detection.

Deep learning (DL) possesses powerful feature learning and representation capabilities, along with short online computation times. Applying DL to the fault and lightning detection in VSC-MTDC grids can avoid setting criterion thresholds. It shows promise in resolving the issues of low detection accuracy and slow detection speed inherent in traditional methods. The application of artificial intelligence algorithms in the field of power system fault detection has become increasingly widespread. Reference [11] proposes a fault detection method based on artificial neural network for multi-terminal DC grids. The fault detection accuracy of this method significantly decreases as fault resistance increases, necessitating further improvements. Reference [12] proposes a fault location method for HVDC transmission lines using Hilbert–Huang transform and convolutional neural network (CNN). The method achieves high localization accuracy and exhibits robustness against varying fault resistance. The authors in reference [13] devise an intelligent fault detection scheme for microgrid based on wavelet transform and deep neural networks. Reference [14] proposes a novel deep CNN transformer model to automatically detect faults for IEEE 14-bus distribution system. The transformer encoder utilizes an attention mechanism to focus on significant time steps to extract the context of the temporal current data. However, none of references [11-14] addressed the identification of internal lightning interference. The idea of using deep neural networks and attention mechanisms to enhance fault and lightning detection accuracy is worth studying.

To address the above challenges, we propose an intelligent fault and lightning detection algorithm for VSC-MTDC grids based on S transform and Residual Network with hybrid attention mechanism (RWHAM). The main contributions of this work are as follows:

(1) The ICTWs of the VSC-MTDC grid are converted to time-frequency matrix by performing S transform, and visualized as the 2D time-frequency image, which well characterize the time-frequency features of double-ended ICTWs.

(2) The RWHAM model is constructed. The key information in the time-frequency image is mined using the hybrid attention mechanism, and the weight of the key information is increased, which improves the fault and lightning detection ability of the RWHAM model.

(3) The 21,620 test datasets obtained through PSCAD/EMTDC simulation proved that the proposed algorithm can achieve reliable and rapid fault and lightning detection for VSC-MTDC grids, and it exhibits excellent generalization ability.

The rest of this paper is organized as follows. Section II introduces the detail information of the research system. Section III describes the process of performing S-transform on ICTWs and visualizing them as images. In section IV, the RWHAM based fault and lightning detection algorithm is presented. Section V validates the effectiveness of the proposed method. Lastly, Section VI gives the conclusion.

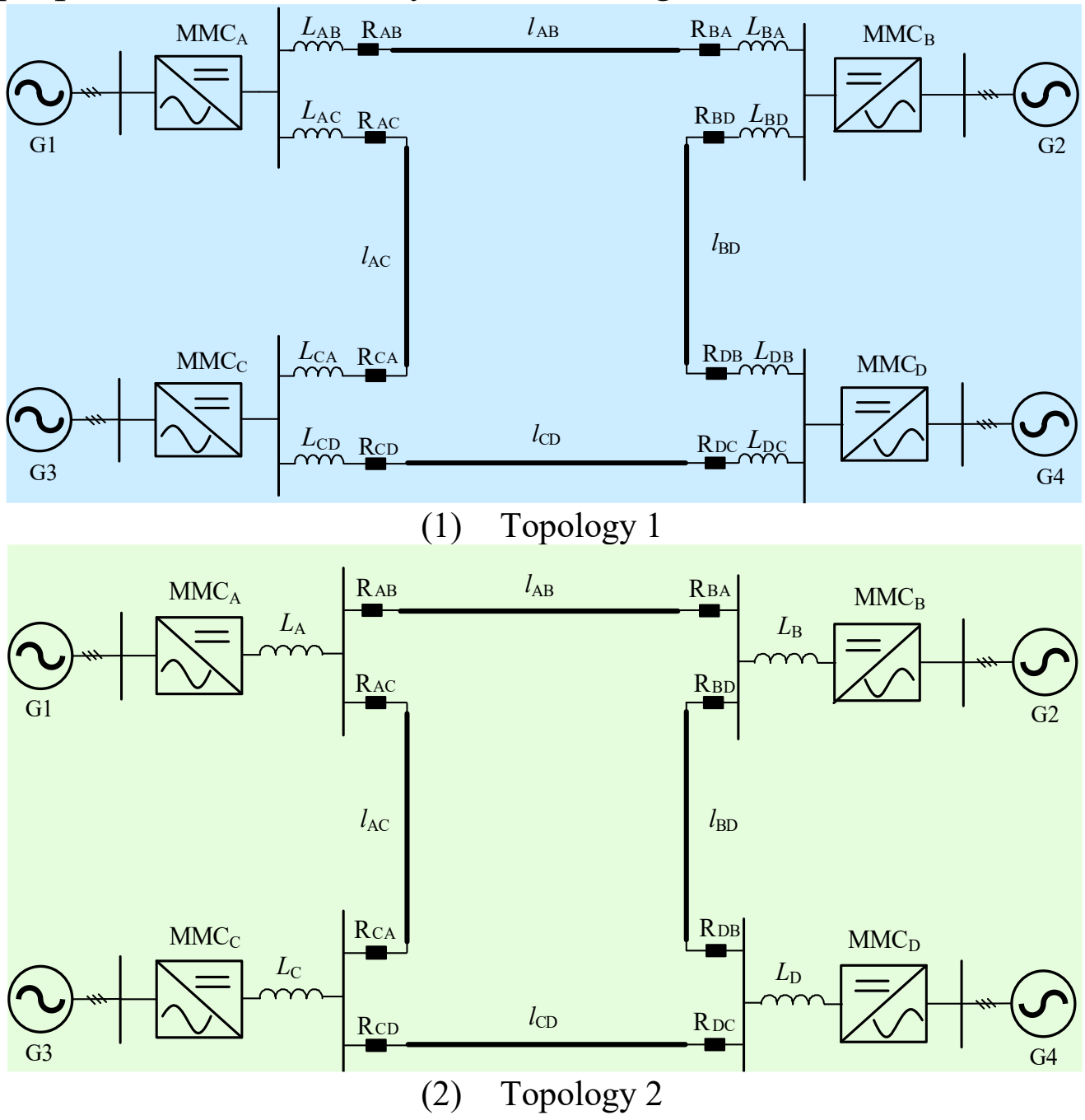

Fig.1 Topologies of the research system

## II. Research System

A four-terminal bipolar loop VSC-HVDC grid is adopted as the test system. Four bipolar converter stations (A, B, C, D) are all composed of the half-bridge modular multilevel converter (MMC). $MMC_A$ adopts constant dc voltage and reactive power control, and the other MMC converter stations adopt constant active and reactive power control. To enhance the stability of the system after fault occurrence, current-limiting reactors $L_{ij}$ or $L_i$ are installed either at both terminals of the DC line or at the outlet of the converter station. Topology 1 and topology 2 are both investigated, as shown in Fig.1 (1) and Fig.1 (2), respectively. The main parameters of the investigated system are listed in Table I. All DC transmission lines $l_{ij}$ (i, j= A, B, C, D) adopt the frequency-dependent overhead line (OHL) model, and its configuration parameters are shown in Fig. 2. The length of all OHLs is 100km. $R_{ij}$ are the relay protections at both terminals of the OHL.

TABLE I
Main Parameters of the Investigated System

| Parameter | Value |
|---|---|
| Rated DC voltage/kV | ±500 |
| Rated AC voltage/kV | 220 |
| Number of sub-modules (SM) per arm $N$ | 200 |
| SM capacitance $C_{SM}$ /mF | 10 |
| Arm inductor $L_{arm}$/mH | 29 |
| Current-limiting reactors $L_{ij}$ or $L_i$ /mH | 200 |
| Rated Capacity/MVA (Station $A$ and Station $B$) | 1500 |
| Rated Capacity/MVA (Station $C$ and Station $D$) | 3000 |

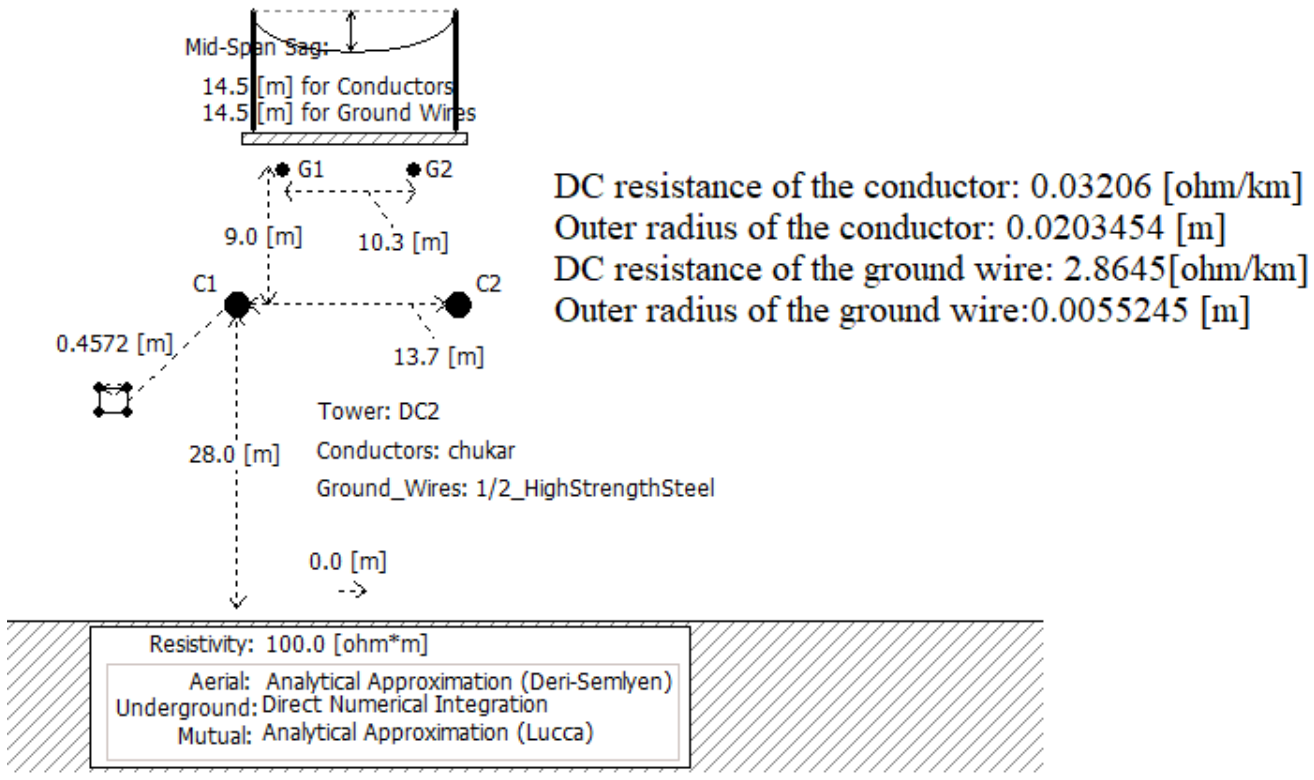

Fig. 2. Configuration of the DC line.

## III. Time-Frequency Feature Extraction using S-Transform

To obtain time-frequency characteristics of wide-band transient signals, discrete S-transform is performed on the ICTWs at both line terminals. Referring to the reasons in [8], we first perform phase-modal transformation on the bipolar DC line currents to decouple them, and perform S-transform on 1-mode current traveling waves. S-transform can achieve higher frequency resolution in low frequency bands and higher time resolution in high frequency bands. S-transform can be expressed as a function of the Fourier transform $H(f)$ of the signal $h(t)$ [9, 15]:

$$S(\tau,f)=\int_{-\infty}^{\infty}H(\alpha+f)e^{-\frac{2\pi^2\alpha^2}{f^2}}e^{i2\pi\alpha\tau}d\alpha \qquad f\neq 0 \tag{1}$$

The discrete representation of S transform is as follows [9, 15],

$$\begin{cases} S\left[jT,\dfrac{n}{NT}\right]=\displaystyle\sum_{m=0}^{N-1}H\left[\dfrac{m+n}{NT}\right]e^{-\frac{2\pi^2m^2}{n^2}}e^{\frac{i2\pi mj}{N}} & n\neq 0 \\ S\left[jT,0\right]=\dfrac{1}{N}\displaystyle\sum_{m=0}^{N-1}h\left[\dfrac{m}{NT}\right] & n=0 \end{cases} \tag{2}$$

where $f$ is equal to $n/NT$; $\tau$ is equal to $jT$; $T$ is the sampling interval; $j,m$=0,1,…,$N$-1; $n$=0,1,…,$N$/2. The collected signal $h[jT]$ with $N$ discrete points is performed S transform based on (2), and a two-dimensional complex time-frequency matrix $\boldsymbol{S}$ can be obtained as follows,

$$\boldsymbol{S}=\begin{bmatrix} S(0,0) & \cdots S(0,b) & \cdots S(0,N\text{-}1) \\ \vdots & \vdots & \vdots \\ S(a,0) & \cdots S(a,b) & \cdots S(a,N\text{-}1) \\ \vdots & \vdots & \vdots \\ S\left(\frac{N}{2},0\right) & \cdots S\left(\frac{N}{2},b\right) & \cdots S\left(\frac{N}{2},N\text{-}1\right) \end{bmatrix} \tag{3}$$

The row vector of the matrix $\boldsymbol{S}$ represents the time-domain characteristics of the signal at a certain frequency, and the column vector represents the frequency-domain characteristics of the signal at a certain moment. The element $\boldsymbol{S}(a, b)$ represents the $b$th sampling point at the $a$th frequency. The frequency of the $a$th row is as follows,

$$f_a=\frac{f_s}{N}a \tag{4}$$

where $f_s$ is the sampling frequency.

The ICTW S-transform window is 0.2ms after the appearance of fault or disturbance signatures. The 0.2 ms time window is selected to ensure accurate extraction of the ICTW while avoiding interference from subsequent reflected and refracted waves. Furthermore, an excessively long window would unnecessarily prolong the detection process. For a time-frequency image from a single terminal of the line, the horizontal axis represents time, ranging from 0 to 0.2 ms, and the vertical axis represents frequency, ranging from 0 to 50 Hz. Then the two generated time-frequency matrices on both terminals are concatenated along horizontal axis and visualized as a 224×224-pixel, 3-channel image using Matlab's ind2rgb function with the Jet colormap.

Taking $R_{AB}$ and $R_{BA}$ of topology 1 as examples, the time-frequency images of internal positive-pole-to-ground faults (PGFs) at different fault locations and with different fault resistances $R_f$ on $line_{AB}$ are shown in Table II. Time-frequency images of external pole-to-pole faults (PPFs) at different fault locations and with different fault resistances on $line_{AC}$ and $line_{BD}$ are shown in Table III. Meanwhile, when the internal 1.2/50us negative-polarity double-exponential lightning interference with current amplitude of 15kA and lightning channel impedance of 300Ω occur at different positions along $line_{AB}$, $line_{BD}$ and $line_{AC}$, the time-frequency images are shown in Table IV, respectively. The different colors in the images represent the different values of the ICTW at varying time instants and frequencies, as indicated by the colorbar. It can be observed that for internal faults and lightning interference, the time-frequency images of ICTWs on both terminals exhibit high similarity in color distribution and variation trends from early to late time periods and from low to high frequencies. In contrast, for external faults and lightning interference, this similarity is significantly lower. The reason is that under internal faults and lightning interference, the overall variation trends of ICTWs on both line terminals remain consistent in time and frequency domains, whereas under external events, these trends become opposite. Moreover, since faults and lightning interference exhibit different distributions in high and low frequencies, their corresponding time-frequency images show distinct color patterns across different frequencies. By extracting the complex features of time-frequency images under various conditions, internal faults, internal lightning interference, as well as external faults or lightning interference can be detected accurately.

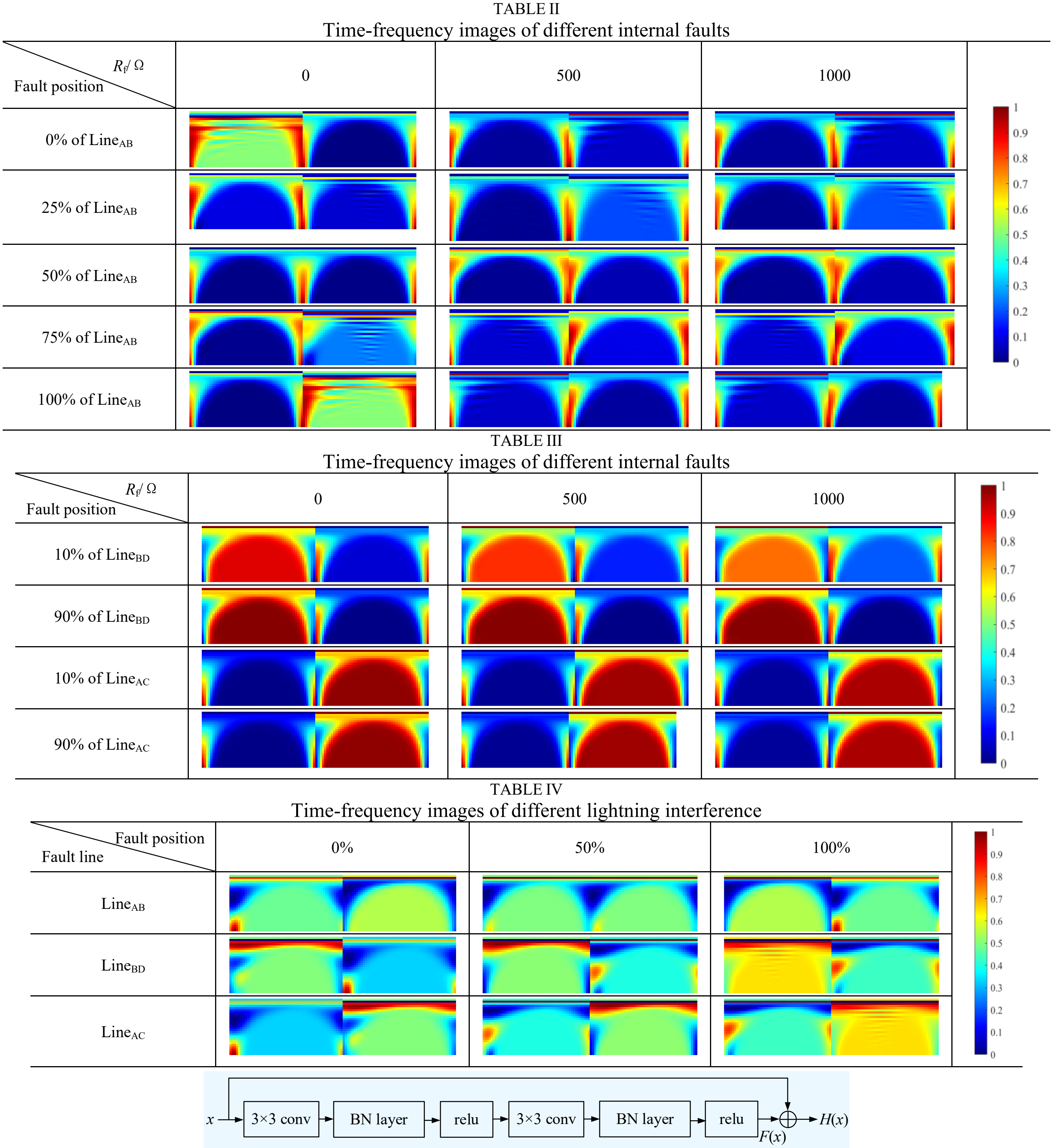

TABLE II
Time-frequency images of different internal faults

| Fault position \ $R_f$/ Ω | 0 | 500 | 1000 |
|---|---|---|---|
| 0% of $Line_{AB}$ | | | |
| 25% of $Line_{AB}$ | | | |
| 50% of $Line_{AB}$ | | | |
| 75% of $Line_{AB}$ | | | |
| 100% of $Line_{AB}$ | | | |

TABLE III
Time-frequency images of different internal faults

| Fault position \ $R_f$/ Ω | 0 | 500 | 1000 |
|---|---|---|---|
| 10% of $Line_{BD}$ | | | |
| 90% of $Line_{BD}$ | | | |
| 10% of $Line_{AC}$ | | | |
| 90% of $Line_{AC}$ | | | |

TABLE IV
Time-frequency images of different lightning interference

| Fault line \ Fault position | 0% | 50% | 100% |
|---|---|---|---|
| $Line_{AB}$ | | | |
| $Line_{BD}$ | | | |
| $Line_{AC}$ | | | |

Fig.3 Architecture of the typical residual block

## IV. RWHAM Based Fault and Lightning Detection Algorithm

To achieve a better trade-off between the speed and reliability of protection in MTDC grids, this paper proposes an intelligent fault and lightning detection algorithm based on RWHAM. The RWHAM model exhibits powerful feature learning and image classification capabilities, enabling accurate extraction of complex time-frequency characteristics from ICTWs in VSC-MTDC grids that are difficult to capture using traditional analytical methods [16-18]. Therefore, by inputting the time-frequency feature map of the ICTW into RWHAM, the DC line internal fault, internal lightning interference, as well as external fault or lightning interference can be identified accurately and quickly.

### A. Resnet structure

ResNet is a classic and stable deep learning architecture, which shows good performance in image classification. The core idea of ResNet is to learn the residual between inputs and outputs, rather than directly learning the target mapping [16-18]. The residual mapping function can be expressed as follows：

$$F(x)=H(x)-x \tag{5}$$

where $x$ is the input quantity; $H(x)$ is the expected output quantity; $F(x)$ is the residual learned by ResNet. The residual block serves as the fundamental building block of ResNet. The architecture of a typical residual block is illustrated in Fig. 5. It consists of two convolutional layers, each followed by a batch normalization (BN) layer and relu activation function. If there is a network degradation problem, the quality of the network can be guaranteed by making $F(x) = 0$. Moreover, optimizing the residual function $F(x)$ is simpler than optimizing the original mapping function, making it easier for ResNet to learn more features from the data. Thus, through residual learning, deep networks can be optimized more effectively. The mathematical model of ResNet can be expressed as follows:

$$y=F\left(x,\{W_{\mathrm{i}}\}\right)+x \tag{6}$$

where $x$ and $y$ are input and output vectors, respectively. $\{W_{\mathrm{i}}\}$ represents the trainable parameters of Convolution blocks, including weights and biases. $F(\cdot)$ represents the residual mapping.

### B. Bottleneck Attention Module

The main structure of Bottleneck Attention Module (BAM) consists of a parallel channel attention module and a spatial attention module [16,17,19]. To focus on the overall importance across channel dimensions, the channel attention module first uses global average pooling operation to aggregate feature maps in each channel, disregarding spatial location variations. Then the channel attention module compresses the channels of the feature maps to $C$/16 via a 3×3 convolution layer, effectively extracting the most discriminative channel features while filtering out redundant information. Subsequently, relu activation function is applied to learn non-linear interdependencies among all channels, thereby enhancing the representational capacity of the network. This is followed by a second 3×3 convolution layer to restore the original channel depth $C$. Finally, sigmoid activation function is applied to generate channel attention weights, where more informative channels are assigned higher values. The channel attention map is shown as follows:

$$M_{\mathrm{c}}(X)=\mathrm{sigmoid}\left(\mathrm{conv}_2^{1\times1}\left(\mathrm{relu}\left(\mathrm{conv}_1^{1\times1}\left(\mathrm{AvgPool}(X)\right)\right)\right)\right) \tag{7}$$

where $X$ is the feature map, and $X\in R^{C\times H\times W}$. The spatial attention module learns weights for each spatial position in the feature maps by aggregating channel information. First, average pooling and max pooling operations are separately performed along the channel dimension, with their results concatenated along channel dimension to produce feature maps of shape [2×$H$×$W$]. Then, a 7×7 convolutional layer with a large receptive field captures spatial contextual relationships in the feature maps, while reducing the output channels to 1. Finally, a sigmoid activation function generates the spatial attention weights, where critical regions are assigned higher weights. The spatial attention map is shown as follows:

$$M_{\mathrm{s}}(X)=\mathrm{sigmoid}\left[\mathrm{conv}_3^{7\times7}\left(\mathrm{maxpool}(X),\mathrm{avgpool}(X)\right)\right] \tag{8}$$

By combining the channel attention module and the spatial attention module, the important feature information in the input data can be better captured and the interference of irrelevant features can be suppressed, which improves the model performance. The output of BAM can be indicated as follows:

$$Y=X+\left(M_c(X)+M_s(X)\right)\cdot X \tag{9}$$

### C. The proposed RWHAM model

In the structure of ResNet, each convolutional layer is responsible for extracting features from the input image. However, not all channels contain semantically meaningful features, as certain channels may primarily contain noise or redundant patterns. Crucially, the most discriminative information tends to be spatially localized within specific receptive fields. For the detection of internal and external faults or lightning interference, the general variation trends of the ICTWs on both line terminals are the most critical information, so the similarity of color distribution and variation trends on both terminals should be given prioritized attention. Moreover, for the detection of internal faults and internal lightning interference, the distinct color distributions across low and high frequencies in the time-frequency images should be emphasized for the learning process. Therefore, it is necessary to incorporate channel and spatial attention mechanisms into feature extraction to increase the sensitivity of the network to important information. Therefore, the hybrid attention mechanism, BAM, is integrated into ResNet, and both spatially salient regions and semantically important channels in the input images are assigned higher weights.

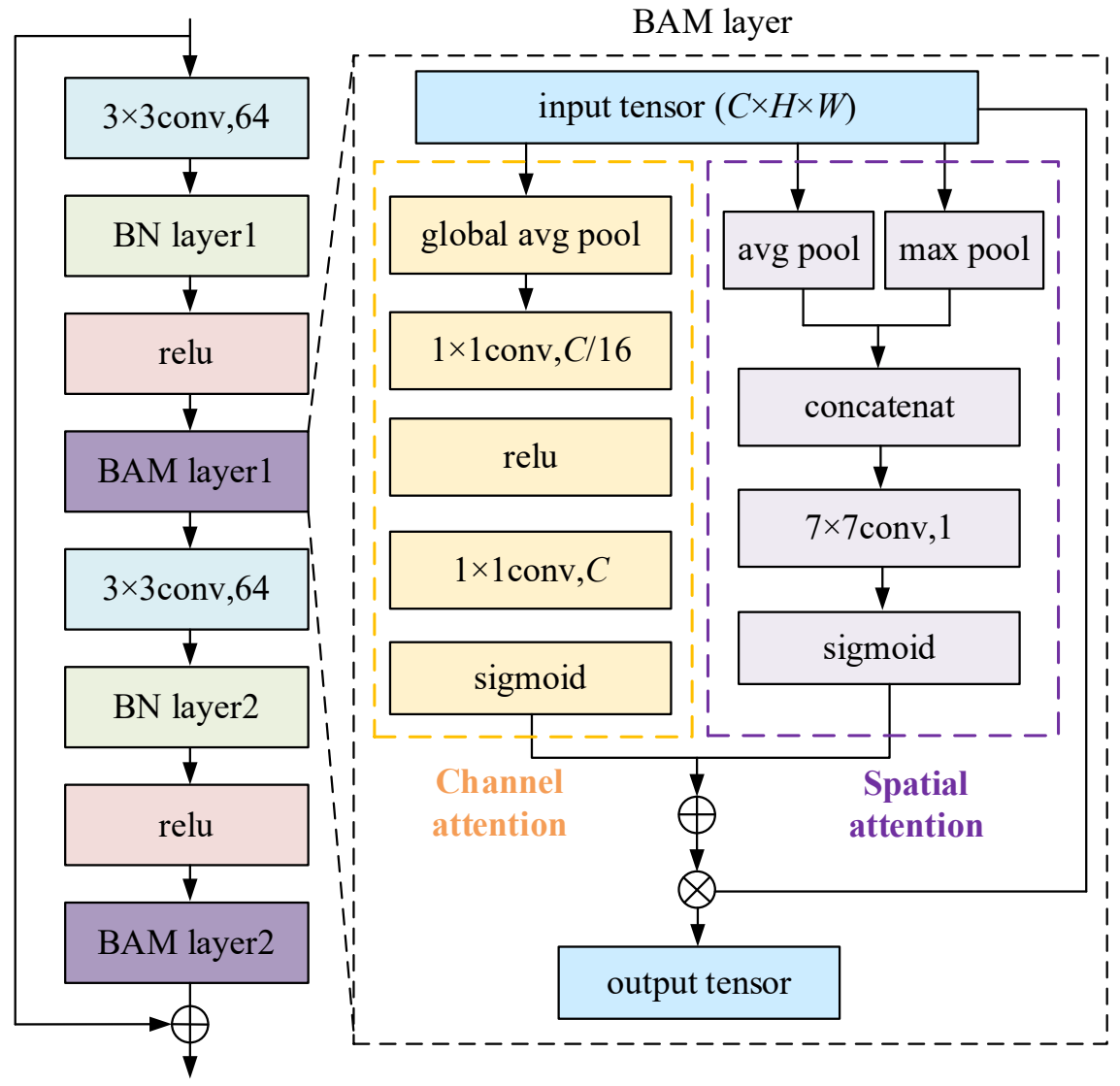

Fig.4 The structure of RB_block

We propose a new convolutional structure named RB_block, which incorporates BAM into the residual block. The structure of RB_block is shown in Fig. 4. In RB_block, a BN layer is appended after each of the two convolutional layers to normalize the feature maps. This reduces the covariate bias present in the original feature maps and makes it easier for the network to converge. A BAM layer is integrated after relu activation function. This layer enables RB_block to properly weight the most informative channels and spatial regions in the feature image. Consequently, the RB_block becomes more sensitive to the similarity of color variation trends in time-frequency images on both line terminals, as well as the different color distributions in low and high frequencies. This heightened sensitivity effectively improves the detection accuracy of line internal and external faults or lightning interference.

This paper introduces the integration of several RB_blocks into ResNet34 architecture, forming a novel RWHAM model, as shown in Fig.5. The RWHAM model initiates with a 3×3 convolutional layer coupled with a BN layer and relu activation function, which performs preliminary extraction of low-level features from the input feature maps. The core structure of RWHAM model comprises four cascaded residual layers with 3, 4, 6 and 3 RB_blocks respectively, where the channel dimensions progressively expand from 64 to 128, 256 and 512. The first residual block in residual layers 2-4 employs a stride of 2, while all other blocks maintain a stride of 1. These four residual layers work hierarchically to extract high-level features from the input feature maps, generating discriminative high-dimensional representations crucial for accurate detection of line internal and external faults or lightning interference. The network concludes with a global average pooling layer that compresses the $512\times H\times W$ feature map generated from the last residual layer into a $512\times 1\times 1$ vector, followed by a fully-connected layer that produces the final classification result. The hyperparameters such as the input and output size of each layer, as well as the filter size of each convolution layer, are optimized using Bayesian optimization implemented in KerasTuner. Specifically, KerasTuner models the relationship between hyperparameters and validation accuracy using a probabilistic surrogate model and iteratively selects the most promising configuration to balance exploration and exploitation [19-21]. The hyperparameters of core layers in BAM, RB_blocks and RWHAM model are elaborated in Appendix A respectively.

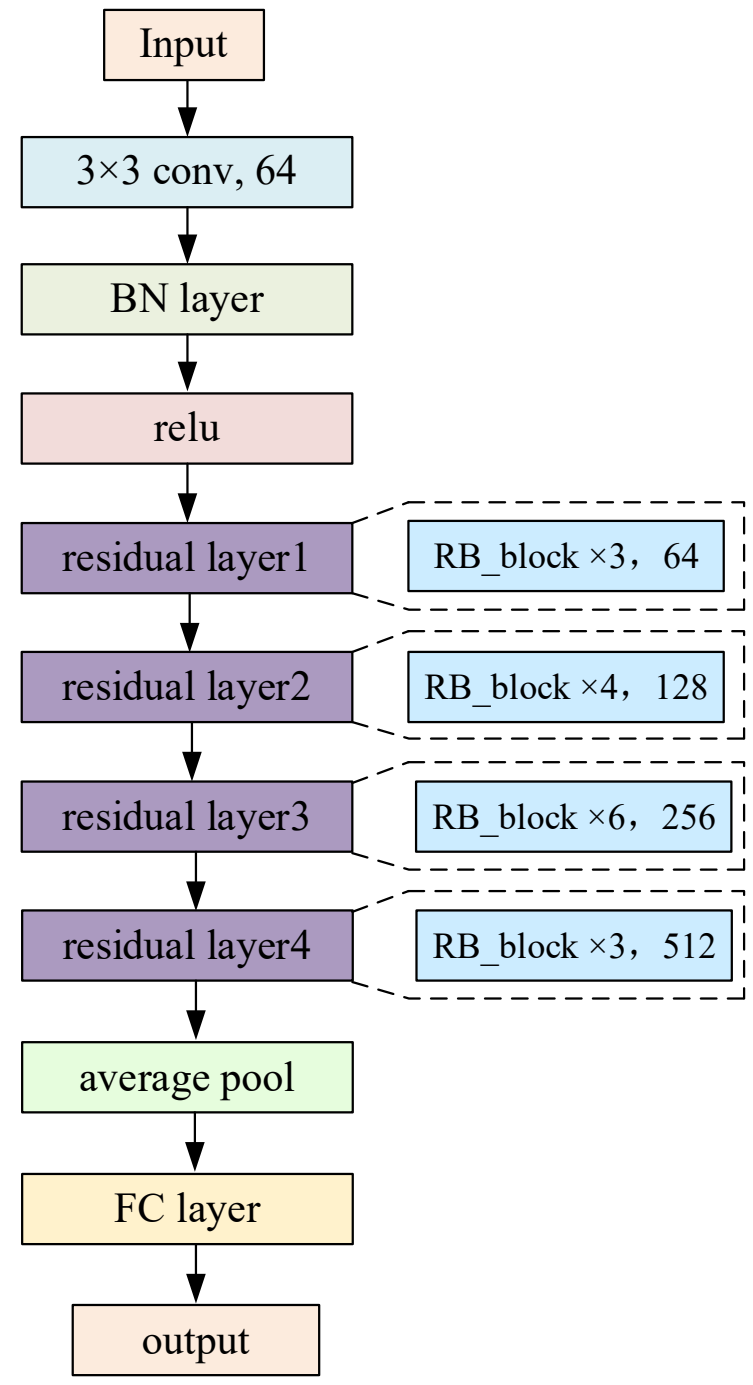

Fig.5 The structure of RWHAM model

The interpretability of the RWHAM model is demonstrated using spatial attention as a representative case. The spatial attention maps output by the RWHAM model under internal fault, external fault, and internal lightning interference are shown in Fig. 6. It can be observed that the RWHAM model generally tends to assign higher attention weights to regions with high amplitudes in the time-frequency images. For this external fault, the amplitude of the left-terminal ICTW is higher, so it is assigned more attention weight. Since the primary difference between the time-frequency images of these internal and external faults lies in the amplitude of the left-terminal ICTW, assigning higher spatial attention weights to the left portion of this external fault time-frequency image enables more accurate internal and external fault detection. Furthermore, regarding internal lightning interference, as ICTWs on both terminals exhibit higher amplitudes in the high-frequency bands, higher attention weight is allocated to the high-frequency regions. In contrast, under internal and external faults, ICTWs on both terminals have higher amplitudes in the low-frequency bands, so higher attention weight is assigned to the low-frequency regions. Given that the intrinsic discrepancy between ICTWs under faults and lightning interference lies in the energy distribution across high and low frequency bands, the RWHAM model can more accurately distinguish lightning interference and faults through this spatial attention mechanism.

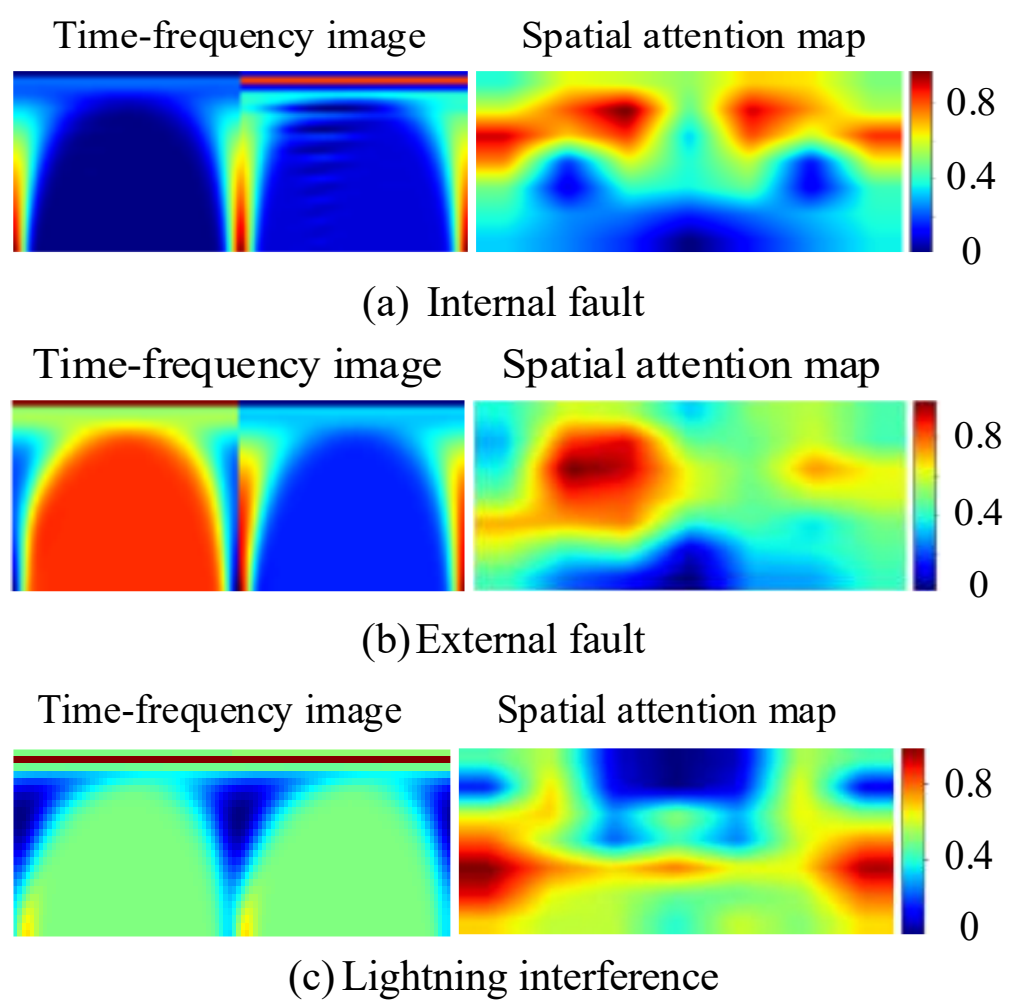

Fig.6 Spatial attention maps under different conditions

### *D. The training process*

Taking $R_{\mathrm{AB}}$ and $R_{\mathrm{BA}}$ as examples, the training datasets containing 21,968 time-frequency images are generated by simulating different fault conditions. Topology 1 and topology 2 are both simulated. The fault and lightning locations include 0% to 100% with step of 5% from initial terminal along all DC lines, as well as all DC buses (A, B, C, D) and AC buses (A, B, C, D). The fault types in the DC area include PGF and PPF. And the fault types in the AC area include AG, AB, ABG and ABCG. For internal faults on $line_{\mathrm{AB}}$, the simulated fault resistances are from 0 Ω to 1000 Ω with step of 10 Ω. For external DC PPFs, the simulated fault resistances are from 0 Ω to 500 Ω with step of 10 Ω. For external DC PGFs and AC faults, the simulated fault resistances are from 0 Ω to 300 Ω with step of 10 Ω. (External faults with fault resistances higher than these values will not trigger the protection.) The simulated lightning interference cases include 1.2/50 us and 8/20 us double-exponential waveforms with amplitudes of -15 kA and -25 kA, respectively [22-24]. The lightning interference occurs on the positive-pole and negative-pole of DC lines and buses, as well as three phases of AC buses, respectively. The lightning channel wave impedance is 300Ω. The simulation frequency is set to 100 kHz.

For the protection of DC lines, the trip signal is only issued when internal faults occur, so it is necessary to distinguish between internal faults and internal lightning interference. However, for external faults and external lightning interference, no trip signal is issued by the protection, so it is unnecessary to distinguish between them. Therefore, during the training and test process, time-frequency images are classified into three categories: internal faults, internal lightning interference, and external faults or lightning interference, labeled as 0, 1 and 2, respectively.

To standardize the images fed into RWHAM, all 21,968 time-frequency images of the training datasets are resized and cropped to uniform dimension of 224×224 pixels. Subsequently, each image is converted into a 3×224×224 PyTorch tensor. To eliminate the dimensional differences of data from different channels and promote faster convergence of the RWHAM model optimizer, as well as make the gradient update more stable, tensor normalization is performed according to the following formulation:

$$x^{'} = \frac{x-\mu}{\sigma} \tag{10}$$

$x^{'}$ and $x$ represent the normalized tensor values and original tensor values of each channel respectively. $\mu$ and $\sigma$ denote the mean and standard deviation of each channel's tensor values. Finally, the normalized tensor data along with their corresponding labels are fed into RWHAM for training.

The loss function $L$ consists of cross-entropy loss term and $L2$ regularization term, commonly used in classification problems, and can be described by the following equation:

$$L(x) = -\frac{1}{N}\sum_{i=1}^{N}\sum_{c=1}^{C} p(x_{i,c})\log\left(q(x_{i,c})\right) + \lambda\sum\omega_j^2 \tag{11}$$

where $N$ is the batch size, which is set to 128. $C$ is the total number of classes. $p(x_{i,c})$ represents the true distribution of the $i$-th sample for class $c$ in one-hot encoding (where only the correct class is 1 and others are 0). $q(x_{i,c})$ represents the predicted probability distribution obtained by RWHAM. To prevent RWHAM from overfitting, the $L2$ regularization term $\lambda\sum\omega_j^2$ is incorporated. $\lambda$ is the regularization term strength, which is set to $1\times10^{-5}$. $\sum\omega_j^2$ is the sum of squares of all weights in RWHAM after the $j$-th iteration. We use the AdamW optimizer to find the optimal values for all the trainable parameters. The learning rate is set to 0.001. After 30 training epochs, the loss value decreased to $2.12\times10^{-5}$. The RWHAM model was constructed and trained on Python 3.12.7 with PyTorch 2.5.1 framework.

The overall architecture of the proposed intelligent fault and lightning detection algorithm is provided in Fig. 7.

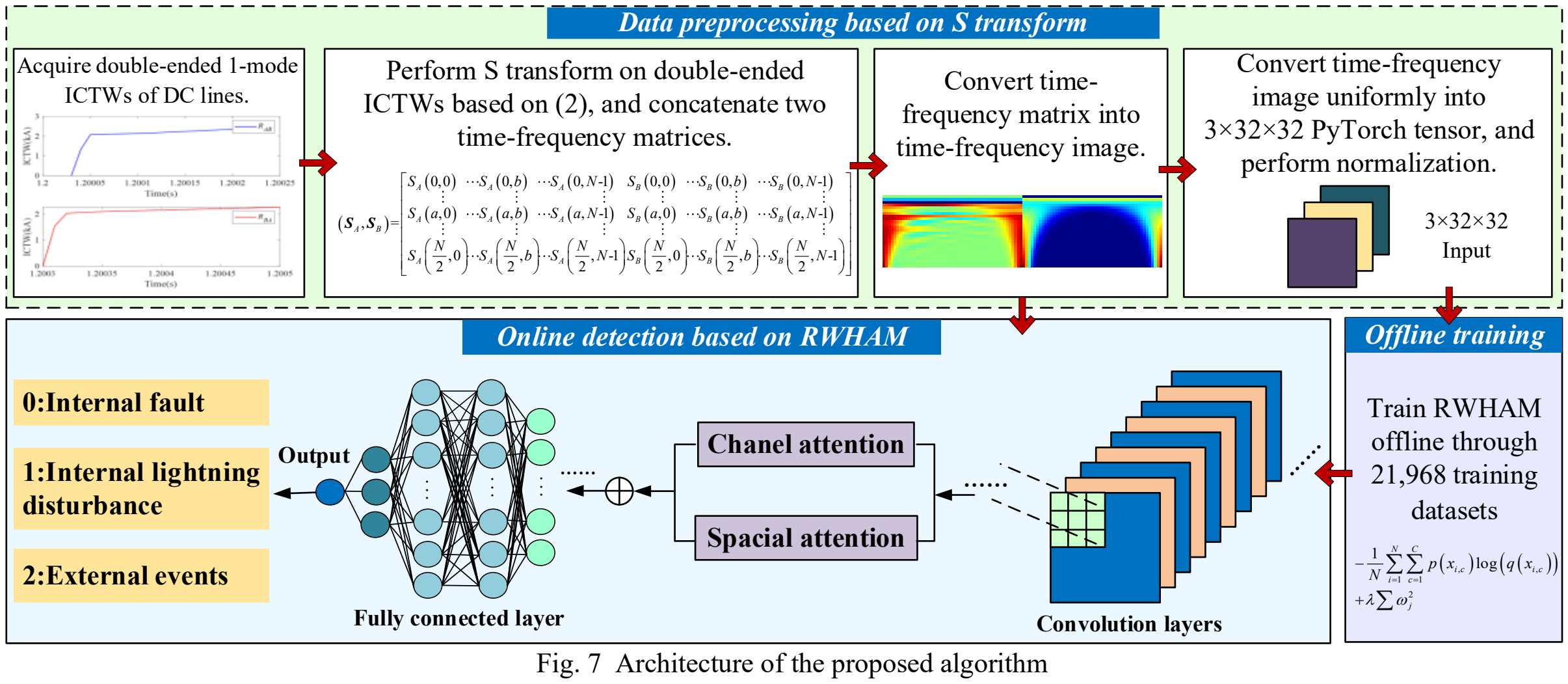

Fig. 7 Architecture of the proposed algorithm

## V. Simulation Validation and Analysis

The detailed model of the VSC-MTDC grid provided in Fig. 1 is developed in PSCAD/EMTDC to validate the superiorities of the proposed intelligent fault and lightning detection algorithm. 21,968 diverse fault cases described in Section IV are simulated to generate the training datasets. All time-series simulations and numerical calculations are conducted on a computer with an Intel Core i7-1260P CPU.

### A. Typical test

To verify the fault and lightning detection algorithm suitable for different fault and lightning locations, types and so on, another 21,620 fault cases different from training datasets are simulated on the research system to generate the test datasets. Details of the configurations for the test datasets are listed in Table V. The test datasets contain 7,600 samples labeled as 0, 1,368 samples labeled as 1, and 12,652 samples labeled as 2.

In order to evaluate the test results more comprehensively, the confusion matrix of the test results was calculated and the results are shown in Fig. 8. The accuracy, precision, recall and F1 score of each label are calculated. Taking label 0 as an example, the calculation methods are shown as (12)-(15), with other labels following the same rule. Where T0, T1 and T2 represent the number of samples with true labels 0, 1 and 2 respectively that are correctly classified. F0, F1 and F2 represent the number of samples with true labels 0, 1 and 2 respectively that are misclassified. $F1_0$ and $F2_0$ represent the number of samples with true labels 1 and 2 respectively that are misclassified as 0. Furthermore, the standard deviation of the accuracy, as well as the average of the accuracy, precision, recall and F1 score of each label across 10 runs are calculated, as shown in Table VI.

TABLE V
Configurations for the Test Datasets

| Parameter | Possible configuration | Count |
|---|---|---|
| Topology | Topology 1 and Topology 2 | 2 |
| Fault and lightning location | Line AB, Line BC, Line AC, Line CD: from 5% to 95% with step of 5%;<br>DC bus A, B, C, D;<br>AC bus A, B, C, D; | 84 |
| Fault type | DC fault: negative-pole-to-ground fault (NGF), PPF.<br>AC fault: (B/C)G, AC, BC, (AC/BC)G, ABCG | 10 |
| Fault resistance | Line AB: from 5Ω to 1005Ω with step of 10Ω;<br>DC PPFs on other fault locations: from 5Ω to 495Ω with step of 10Ω;<br>DC NGFs and AC faults: from 5Ω to 295Ω with step of 10Ω. | 101 |
| Lightning channel wave impedance | 200Ω, 400Ω and 500Ω | 3 |
| Lightning current amplitude | -5kA, -10kA and -20kA | 3 |
| Lightning exponent type | 2.6/50us, 8/20us | 2 |
| Lightning pole or phase type | DC area: positive pole, negative pole.<br>AC area: A/B/C. | 5 |

$$\text{Accuracy}=\frac{\text{T0+T1+T2}}{\text{T0+T1+T2+F0+F1+F2}} \tag{12}$$

$$\text{Precision}=\frac{\text{T0}}{\text{T0+F0}} \tag{13}$$

$$\text{Recall}=\frac{\text{T0}}{\text{T0+F1}_0\text{+F2}_0} \tag{14}$$

$$\text{F1-score}=\frac{2\times\text{Precision}\times\text{Recall}}{\text{Precision+Recall}} \tag{15}$$

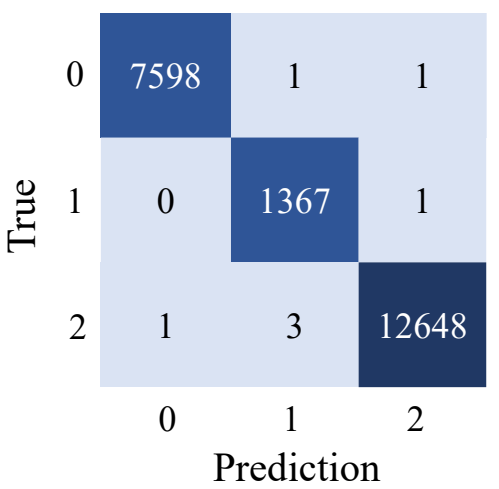

Fig. 8 The confusion matrix of the typical test

The test results demonstrate that the proposed intelligent fault and lightning detection algorithm achieves high average accuracy, precision, recall and F1-score, which are all not below 99.72% for each label across 10 runs. And the standard deviations of accuracy are all less than 0.01%. Therefore, the proposed algorithm can accurately identify various types of internal faults, internal lightning interference and external faults or lightning interference at different locations in VSC-MTDC grids, withstanding fault resistances up to 1005 Ω . And it can be adapted to two topologies without relying on line boundary components. Besides, the proposed algorithm requires no manual setting or adjustment of protection thresholds. Though the powerful learning and classification capabilities of RWHAM for complex features, it can be effectively applied to all the aforementioned different conditions.

TABLE VI
The Typical Test Results for Each Label

| Label | Standard deviation | Accuracy | Precision | Recall | F1-score |
|---|---|---|---|---|---|
| 0 | 0 | 99.97% | 99.97% | 99.99% | 99.98% |
| 1 | 0.01% | 99.97% | 99.92% | 99.72% | 99.82% |
| 2 | 0 | 99.97% | 99.97% | 99.98% | 99.98% |

## B. Generalization ability

The generalization ability of the proposed algorithm is analyzed from the following three typical aspects.

(1) Different voltage level or transmitted power

When the voltage level or transmission power of the VSC-MTDC system varies, the system's impedance parameters remain constant while the line voltages or currents change. This is mathematically equivalent to scaling $x$ to $kx$ in (10), where $k$ represents the voltage and current scaling factor. Under such conditions, both $\mu$ and $\sigma$ scale proportionally to $k\mu$ and $k\sigma$ respectively, while $x'$ maintains its original value. The normalization process effectively standardizes time-frequency images under different voltage levels or transmission powers. Notably, the time-frequency characteristics of line ICTWs learned by the RWHAM model are solely determined by the impedance parameters of the VSC-MTDC grid. Therefore, the trained RWHAM model can be applied to VSC-MTDC grids with different voltage levels or transmission powers.

(2) Different DC lines

To evaluate the algorithm's generalization capability across different DC line types and parameters, the DC OHLs in the research system are replaced with DC cable lines, the configurations of which are shown in Fig. 9 [27]. Additionally, the line lengths were modified to 50 km, 400 km and 800 km respectively. For each DC line type, we re-simulated all 21,620 cases specified in Table I to create new test datasets. The confusion matrices of the test results with different DC lines are shown in Fig. 10. The standard deviation of the accuracy, as well as the average of the accuracy, precision, recall and F1 score of each label across 10 runs are calculated, which are presented in Table VII.

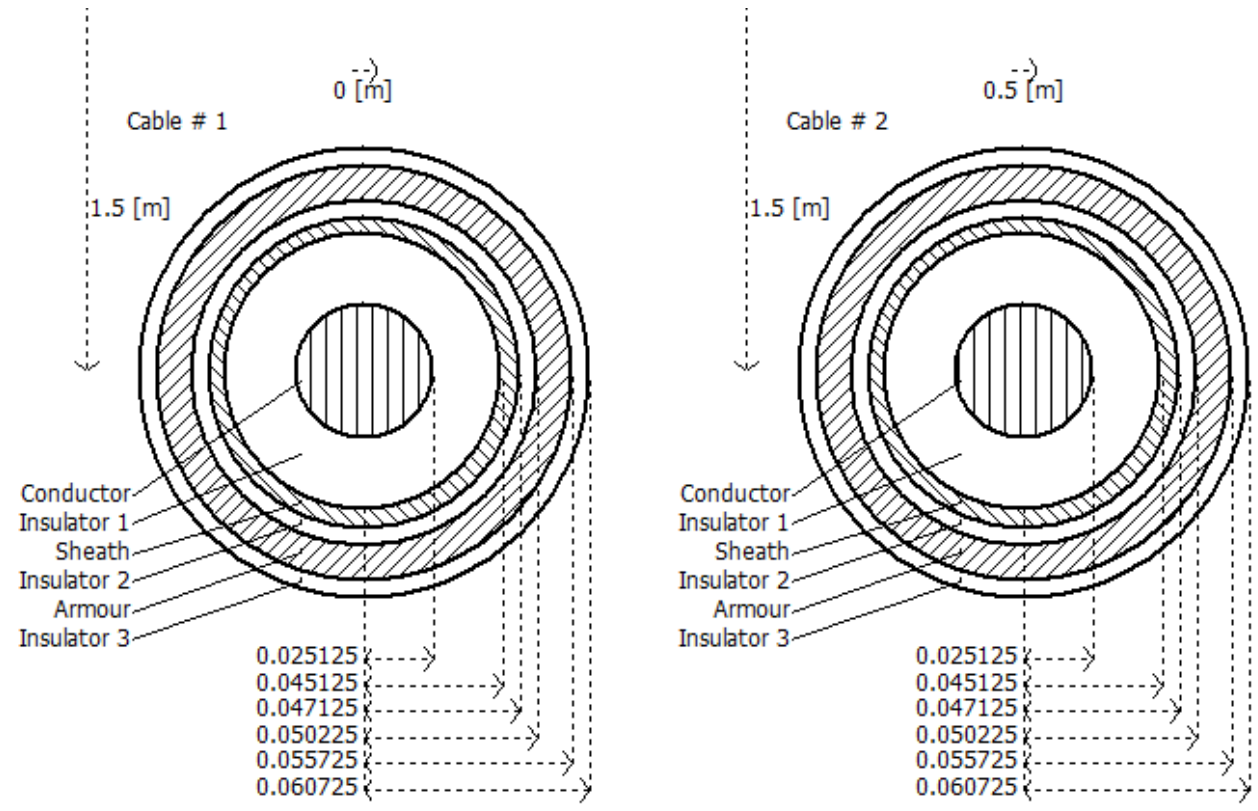

Fig. 9 The configuration parameters of the DC cable line

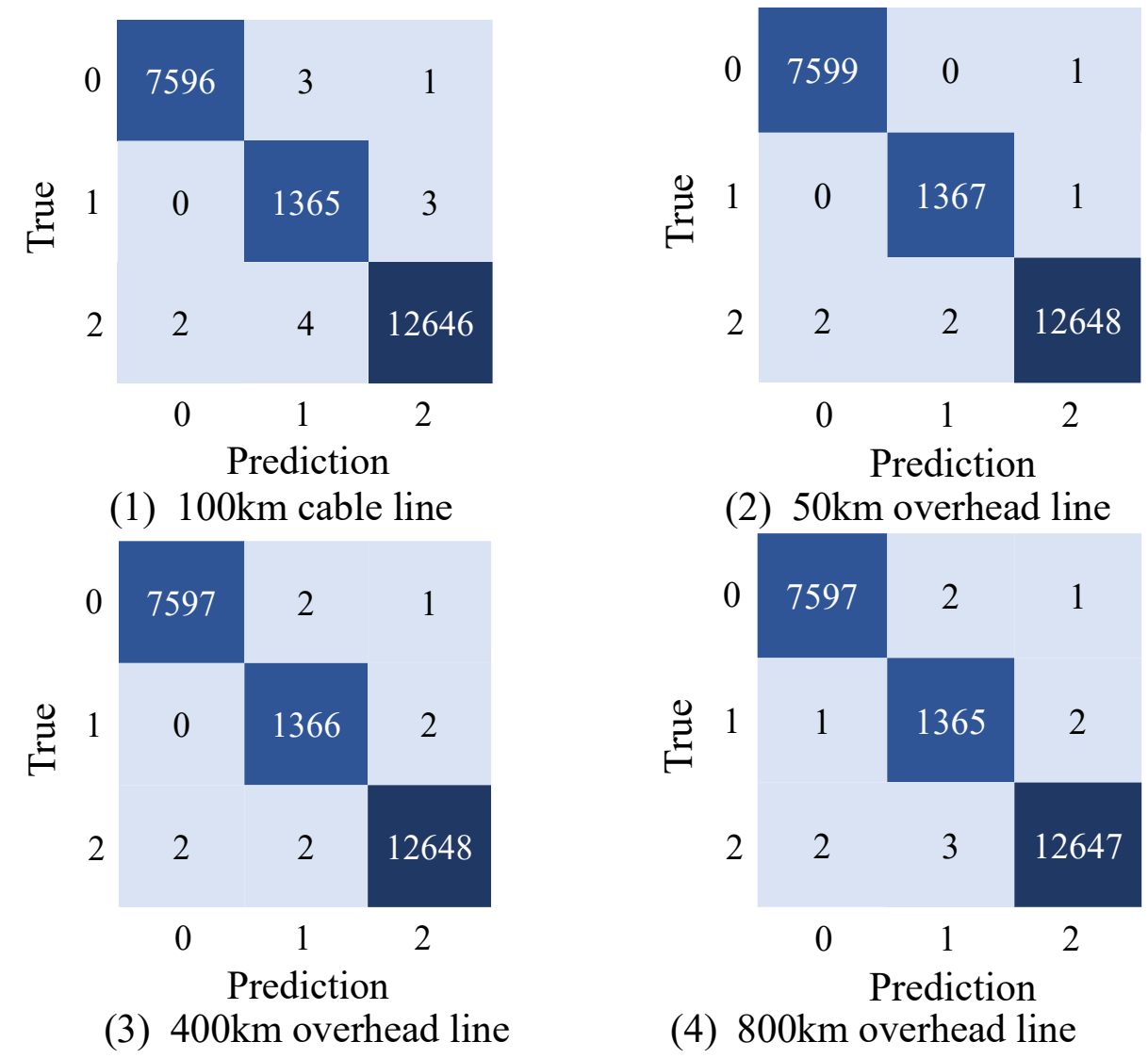

Fig. 10 The confusion matrices of the tests with different DC lines

TABLE VII
Test Results with Different DC Lines

| DC line | Label | Standard deviation | Accuracy | Precision | Recall | F1-score |
|---|---|---|---|---|---|---|
| 100km cable line | 0 | 0.01% | 99.94% | 99.95% | 99.96% | 99.96% |
| | 1 | 0.02% | 99.93% | 99.77% | 99.50% | 99.62% |
| | 2 | 0.01% | 99.94% | 99.95% | 99.97% | 99.96% |
| 50km overhead line | 0 | 0 | 99.97% | 99.99% | 99.97% | 99.98% |
| | 1 | 0.01% | 99.97% | 99.93% | 99.86% | 99.90% |
| | 2 | 0.01% | 99.97% | 99.97% | 99.97% | 99.97% |
| 400km overhead line | 0 | 0.01% | 99.96% | 99.96% | 99.97% | 99.97% |
| | 1 | 0.02% | 99.95% | 99.86% | 99.69% | 99.77% |
| | 2 | 0.01% | 99.96% | 99.97% | 99.97% | 99.97% |
| 800km overhead line | 0 | 0.01% | 99.96% | 99.97% | 99.96% | 99.97% |
| | 1 | 0.02% | 99.94% | 99.77% | 99.66% | 99.72% |
| | 2 | 0.02% | 99.95% | 99.96% | 99.98% | 99.97% |

From Fig.10 and Table VII, it can be seen that when the DC line is changed to cable, the average accuracy, precision, recall and F1-score of the test datasets across 10 runs are all not below 99.50%. Additionally, the standard deviations of the accuracy are less than 0.02%. Consequently, the proposed fault and lightning detection algorithm can also accurately identify different types of internal and external faults as well as lightning interference with DC cable lines. Furthermore, when the line length varies between 50km and 800km, the accuracy, precision, recall and F1-score of the test datasets are all not below 99.64%. Thus, the test results verify that the proposed fault and lightning detection algorithm is nearly immune to line length variations while maintaining excellent generalization ability.

(3) Different current limiting reactors

To verify the generalization capability of the fault and lightning detection algorithm for different current-limiting reactor values, the current-limiting reactor values were adjusted to 50 mH and 500 mH, respectively. For each current-limiting reactor value, the 21,620 faults listed in Table II were re-simulated to form new test datasets. The test results are shown in Table VIII.

From Table VIII, it can be seen that when the current limiting reactors are changed to 50mh and 500mH, the average accuracy, precision, recall and F1-score of the test datasets across 10 runs are all not below 99.72%. Moreover, the standard deviations of the accuracy are all less than 0.02%. Thus, the proposed fault and lightning detection algorithm is almost unaffected by the current-limiting reactor value, and can reliably identify different types of faults and lightning interference on diverse topologies, almost all fault locations with different current limiting reactor values. Its generalization ability for different current limiting reactor is strong.

TABLE VIII
Test Results under Different Current-limiting Reactors

| Current-limiting reactors | Label | Standard deviation | Accuracy | Precision | Recall | F1-score |
|---|---|---|---|---|---|---|
| 50mH | 0 | 0.01% | 99.96% | 99.96% | 99.98% | 99.97% |
| | 1 | 0.02% | 99.95% | 99.91% | 99.72% | 99.81% |
| | 2 | 0.01% | 99.96% | 99.95% | 99.98% | 99.97% |
| 500mH | 0 | 0.01% | 99.97% | 99.97% | 99.96% | 99.97% |
| | 1 | 0.01% | 99.96% | 99.94% | 99.86% | 99.90% |
| | 2 | 0 | 99.97% | 99.97% | 99.97% | 99.97% |

### C. Detection time

$T_D$ is defined as the time period that from the fault or lightning occurrence to the ending of detection, and $T_D$ can be calculated as the following equation,

$$T_D = T_1 + T_2 + T_3 \quad (16)$$

where $T_1$ is the time window of the extracted ICTW, i.e. 0.2ms. $T_2$ includes the travelling wave transmission time $T_2^a$ and the signal propagation time $T_2^b$. $T_3$ comprises the sampling data processing time, and a certain length of margin. The average value of $T_3$ for all of the test datasets is 0.58ms. Considering the maximum detection time of the main protection of VSC-MTDC grids is 5ms, the maximum allowable value of $T_2$ is 4.22ms. The velocity of 1-mode travelling wave in the DC line is 294 km/ms, and the time delay for an optical fiber communication system is around 4.9 us/km. It can be calculated that both $T_2^a$ and $T_2^b$ increase with the DC line length, and $T_2^a$ reaches maximum value when the fault occurs at the line terminal. It can be calculated that when the length of the DC line does not exceed 508km, this algorithm can be used as main protection of VSC-MTDC grids. If the DC line length exceeds 508km, this algorithm can be used as the backup protection of VSC-MTDC grids.

### D. Comparison with other algorithms

To further demonstrate the superiority of the proposed fault and lightning detection algorithm, it is compared with the algorithms proposed in references [4], [5], [9], based on CNN and ResNet in the following four aspects: the accuracy $A_1$ of detecting internal faults, external faults and external lightning interference; the accuracy $A_2$ of detecting internal and external faults (If lightning interference detection capability is not available); the maximum online detection time $T_D$; and whether the protection threshold needs to be set. Both $A_1$ and $A_2$ are calculated based on label 0. Taking three scenarios as examples: the original test system, adjusting the length of the DC OHL to 800 km, and adjusting the current-limiting reactance value to 50mH. The test dataset constructed from 21,620 different cases as described in Table V was used to evaluate different algorithms. The test results are shown in Table IX, where "/" indicates the algorithm lacks the capability to detect lightning interference.

Since the algorithm proposed in reference [4] cannot accurately detect high-impedance internal and external faults,

$A_1$ of it is observably low. Besides, the algorithm proposed in reference [4] relies on the boundary effect of current-limiting reactors and is not applicable to Topology 2 of the test system, so $A_1$ of it is also significantly low. The $A_2$ values of the algorithms proposed in references [5], [9], based on CNN and ResNet have been improved. However, $T_D$ of the algorithm proposed in reference [5] is too long to satisfy the speed requirement for main protection in VSC-MTDC grid. Besides, the algorithms proposed in references [4], [5] and [9] are all based on the fault mechanism of physical models, and all require setting protection thresholds. When the length of the DC overhead line or the current-limiting reactance value changes, their detection accuracy will significantly decrease, necessitating manual threshold adjustments to maintain detection accuracy. Additionally, the generalization capability of CNN and ResNet is limited. When the line length or the value of the current-limiting reactance changes, $A_1$ and $A_2$ exhibit a more significant decline. Among all the algorithms mentioned above, the proposed intelligent fault and lightning detection algorithm achieves the highest detection accuracy. It requires no manual threshold setting and maintains strong generalization capability even when system parameters change. Furthermore, when the DC line length is 100 km, the maximum detection time required by the proposed algorithm is only 2.52 ms, making it suitable as main protection. For the 800 km DC line, due to traveling wave and signal propagation delays, the maximum detection time increases to 8.33 ms, allowing it to serve as backup protection.

TABLE IX

The results of comparison with other algorithms

| Parameter | Algorithm | $A_1$ | $A_2$ | $T_D$/ms | Threshold |
|---|---|---|---|---|---|
| Original system | [4] | / | 59.12% | 2.11 | Yes |
| | [5] | / | 57.17% | 7.23 | Yes |
| | [9] | 95.32% | 98.62% | 2.61 | Yes |
| | CNN | 97.54% | 98.46% | 2.38 | No |
| | ResNet | 98.28% | 99.12% | 2.43 | No |
| | The proposed | 99.97% | 100% | 2.52 | No |
| 800km OHL | [4] | / | 40.12% | 2.11 | Yes |
| | [5] | / | 52.02% | 7.23 | Yes |
| | [9] | 77.23% | 84.10% | 8.37 | Yes |
| | CNN | 94.67% | 96.57% | 8.22 | No |
| | ResNet | 95.48% | 97.04% | 8.27 | No |
| | The proposed | 99.95% | 99.98% | 8.33 | No |
| 50mH current-limiting reactance | [4] | / | 57.08% | 2.11 | Yes |
| | [5] | / | 46.17% | 7.23 | Yes |
| | [9] | 95.04% | 98.18% | 2.61 | Yes |
| | CNN | 96.85% | 98.62% | 2.38 | No |
| | ResNet | 97.33% | 98.70% | 2.43 | No |
| | The proposed | 99.96% | 99.99% | 2.52 | No |

## VI. CONCLUSION

This paper proposes an intelligent fault and lightning detection algorithm for VSC-MTDC grids based on S transform and RWHAM model. Firstly, the current time series data is transformed into 2D images by performing S transform. These images distinctly reveal the time-frequency features of the double-ended ICTWs. Subsequently, the RWHAM model is developed to enhance the weights of critical information within the input images, thereby demonstrating superior feature extraction capability. Comprehensive testing on a dataset of 21,620 cases obtained by PSCAD/EMTDC simulations demonstrates that under nominal grid configurations, the algorithm achieves 99.97% detection accuracy. It effectively distinguishes various lightning interferences, withstands fault resistance up to 1005Ω, operates independently of line boundary components, and provides full-line protection coverage. Furthermore, even when grid parameters vary without model retraining, the fault and lightning detection accuracy remains above 99.94%, showcasing exceptional robustness and generalization. When the DC line length does not exceed 508km, the fault and lightning detection time is within 5ms, fulfilling the stringent speed requirements for primary protection in VSC-MTDC grids. While in other scenarios, it serves as a reliable backup protection strategy.

## APPENDIX A

Table. AI

The hyperparameters of the core layers in BAM

| Layer | | Input size | Filter size | Stride | Output size |
|---|---|---|---|---|---|
| Channel Attention | conv1 | $C \times H \times W$ | 1×1 | 1 | $(C/16) \times H \times W$ |
| | conv2 | $(C/16) \times H \times W$ | 1×1 | 1 | $C \times H \times W$ |
| Channel Attention | conv | $C \times H \times W$ | 7×7 | 1 | $1 \times H \times W$ |

Table. AII

The hyperparameters of the core layers in RB_block

| Layer | Input size | Filter size | Stride | Output size |
|---|---|---|---|---|
| conv1 | $C \times H \times W$ | 1×1 | 1 | $(C/16) \times H \times W$ |
| conv2 | $C \times H \times W$ | 7×7 | 1 | $1 \times H \times W$ |

Table. AIII

The hyperparameters of the core layers in RWHAM

| Layer | Input size | Filter size | Stride size | Output size | Number of RB_block |
|---|---|---|---|---|---|
| conv | $C \times H \times W$ | 3×3 | 1 | $64 \times H \times W$ | / |
| residual1 | $64 \times H \times W$ | / | 2 | $64 \times H \times W$ | 3 |
| residual2 | $64 \times H \times W$ | / | 1 | $128 \times H \times W$ | 4 |
| residual3 | $128 \times H \times W$ | / | 1 | $256 \times H \times W$ | 6 |
| residual4 | $256 \times H \times W$ | / | 1 | $512 \times H \times W$ | 3 |
| average pool | $512 \times H \times W$ | / | / | 512×1×1 | / |
| fully-connected | 512×1×1 | / | / | 1×1×1 | / |

## REFERENCES

[1] Cheng H, Li C, Yang Y, et al. A New Perspective for Frequency and DC Voltage Stability Comprehension in Two-Terminal VSC-HVDC Systems With Virtual Inertia Control," *IEEE Transactions on Industrial Electronics*, vol. 72, no. 3, pp. 3233-3240, March 2025.

[2] Li Q, Wu L, and Wang X, et al. "Comparative analysis of single-machine equivalent methods for heterogeneous PMSG-based wind farm with the VSC-HVDC system,"*Applied Energy*, vol.38, no.124988, 2025,

[3] Xu Z, Li G , Li X, et al. "Transient Stability Analysis of Renewable Power Generations via VSC-HVDC," *IEEE Transactions on Industrial Electronics*, vol. 72, no. 5, pp. 4889-4899, May 2025.

[4] Li Z, Duan J, Lu W, et al. "A Fast Pilot Protection for DC Distribution Networks Considering the Whole Fault Process," *IEEE Transactions on Power Delivery*, vol.37, no.4, pp.3121-3132, August 2022.

[5] Kong F, Hao Z G, Zhang B H. "A Novel Traveling-Wave-Based Main Protection Scheme for ±800 kV UHVDC Bipolar Transmission Lines," *IEEE Transactions on Power Delivery*, vol.31, no.5, pp.2159-2168, October 2016.

[6] Li Y J, Wu L, Li J P, et al. "DC Fault Detection in MTDC Systems Based on Transient High Frequency of Current," *IEEE Transactions on Power Delivery*, vol.34, no.3, pp. 950-962, June 2019.

[7] Zhang C, Li Y, Song G, et al. “Fast and Sensitive Nonunit Protection Method for HVDC Grids Using Levenberg–Marquardt Algorithm,” *IEEE Transactions on Industrial Electronics*, vol.69, no.9, pp.9064-9074, October 2021.

[8] Zhang Y Q, Cong W. “An improved single-ended frequency-domain-based fault detection scheme for MMC-HVDC transmission lines,” *International Journal of Electrical Power & Energy Systems*, vol.125, no.106463, February 2021.

[9] Zhang Y Q, Wang C B, Yu Y. “A pilot protection method based on the similarity of initial current traveling wave time-frequency matrix for VSC-HVDC grids,” *Electric Power Systems Research*, vol.217, no.109116, April 2023.

[10] Alireza P, Hossein I E, Sattar B, et al. “A Fault Detection Algorithm Based on Artificial Neural Network Threshold Selection in Multi-Terminal DC Grids,” *IEEE Transactions on Power Delivery*, vol.38, no.4, pp.2510-2520.

[11] Mou F G, Nien C Y, Wei F C. “Deep-Learning-Based Fault Classification Using Hilbert–Huang Transform and Convolutional Neural Network in Power Distribution Systems,” *IEEE Sensors Journal*, vol.19, no.16, pp.6905-6913, August 2019.

[12] James J. Q. Yu, Hou Y H, Albert Y. S. Lam. “Intelligent Fault Detection Scheme for Microgrids With Wavelet-Based Deep Neural Networks,” *IEEE Transactions on Smart Grid*, vol.10, no.2, pp.1694-1703, March 2019.

[13] Jibin B. T, Saurabh G. C, Shihabudheen K. V. “CNN-Based Transformer Model for Fault Detection in Power System Networks,” *IEEE Transactions on Instrumentation and Measurement*, vol.72, no.2504210, 2023.

[14] Zhang Y, Cong W, Li G, et al. “Single-ended MMC-MTDC line protection based on dual-frequency amplitude ratio of traveling wave,” *Electric Power Systems Research*, vol.189, no.106808, December 2020.

[15] R. G. Stockwell L M, and R. P. Lowe. “Localization of the Complex Spectrum,” *IEEE Transactions on Signal Processing*, vol.44, no.4, pp.998-1001, August 1996.

[16] Song X, Dai Y, Zhou D, et al. “Channel Attention based Iterative Residual Learning for Depth Map Super-Resolution”, *2020 IEEE/CVF Conference on Computer Vision and Pattern Recognition (CVPR).*

[17] Jie L, Wenjie Z, Yuting T, et al. “Residual Feature Aggregation Network for Image Super-Resolution,” *2020 IEEE/CVF Conference on Computer Vision and Pattern Recognition (CVPR)*.

[18] Yin L, Ge W. “Mobileception-ResNet for transient stability prediction of novel power systems,” *Energy*, vol.309, pp.133163, 2024.

[19] Song X, Tan Y, Pang X, et al. “Spatial and channel enhanced self-attention network for efficient single image super-resolution,” *Neurocomputing*, vol.193, no.108928, 2025.

[20] Xuan D.J.N, Liu Y.A. “Methodology for hyperparameter tuning of deep neural networks for efffcient and accurate molecular property prediction,” *Computers and chemical engineering*, vol.193, no.108928, 2025.

[21] Ashley L, Dennis D and Stephen R. N. “Importance of Hyper-Parameter Optimization During Training of Physics-Informed Deep Learning Networks,” *Integrating Materials and Manufacturing Innovation*, vol.14, pp.115–135, 2025.

[22] B. Kermani , R. Shariatinasab, M. Khorshidi, et al. “Risk Analysis of the Lightning-Related Transients on Photovoltaic Systems: Application to a Solar Power Plant Without a Lightning Protection System,”, *IEEE Transactions on Power Delivery*, vol.40, no.1, pp. 618-629, February 2025.

[23] Mohammad J N, Omid H, Arsalan N, et al. “Analyzing the Effect of Lightning Channel Impedance on the Lightning Overvoltages in Wind Turbines,” *IEEE Transactions on Industry Applications*, vol.59, no.5, pp.5352-5362, October 2023.

[24] Yang P, Chen S, and He J, et al. “Lightning Impulse Corona Characteristic of 1000-kV UHV Transmission Lines and Its Influences on Lightning Overvoltage Analysis Results,” *IEEE Transactions on Power Delivery*, vol.28, no.4, pp.2518-2525, October 2013.

[25] Zhang Y, Yu Y, Yang G. “An ultra-fast MMC-HVDC fault location algorithm based on transient voltage features and regression neural network,” *International Journal of Electrical Power and Energy Systems*, vol.162, no.110249, 2024.